\documentclass{epl}
\usepackage{graphicx}
\usepackage{bm}

 \newcommand{\BF}[1]{\mbox{\boldmath $#1$}}
 
\def\deltab{{\BF \Delta}}
\newcommand{\ve}[1]{\bm{\mathbf{#1}}}
\def\comment#1{}

\title{Phase structure of Abelian Chern-Simons gauge theories}
\shorttitle{Abelian Chern-Simons gauge theories}
\author{E. Sm{\o}rgrav\inst{1}, J. Smiseth\inst{1}, A. Sudb{\o}\inst{1} \and F. S. Nogueira\inst{2}}
\institute{\inst{1} Department of Physics, Norwegian University of
Science and Technology, N-7491 Trondheim, Norway\\
\inst{2} Institut f{\"u}r Theoretische Physik,
Freie Universit{\"a}t Berlin,
D-14195 Berlin, Germany}

\pacs{11.15.Ha}{Lattice gauge theory}
\pacs{74.20.-z}{Theories and models of superconducting state}
\pacs{05.30.Pr}{Fractional statistics systems}

\begin{document}

\maketitle

\begin{abstract}
We study the effect of a Chern-Simons (CS) term
in the phase structure of two different Abelian gauge theories.
For the compact Maxwell-Chern-Simons theory, with the
CS-term properly defined, we obtain that for values
$g=n/2\pi$  of the CS coupling with $n=\pm 1,\pm 2$, the
theory is equivalent to a gas of closed loops with contact interaction,
exhibiting a phase transition in the $3dXY$ universality class.
We also employ Monte Carlo simulations to study
the noncompact $U(1)$ Abelian Higgs model with a CS term. Finite size
scaling of the third moment of the action yields critical exponents
$\alpha$ and $\nu$ that vary  continuously with the strength of the
CS term, and a comparison with available analytical results is made.

\end{abstract}


Gauge theories in $2+1$ dimensions are frequently proposed as effective theories
of strongly correlated electron systems in two spatial dimensions and zero
temperature. Strong local constraints on the dynamics of lattice fermion
systems are enforced by fluctuating compact gauge fields which exhibit
topological defects in the form of space-time instantons \cite{polyakov}.
It is conceivable that such effective gauge theories may feature phase
transitions from a phase of bound instantons (deconfined phase) to
a phase of unbound free instantons (confined phase). Such theories
have been proposed as effective theories for high-temperature
cuprate superconductors, chiral spin liquids, and Mott insulators
\cite{Nagaosa,Fradkin:1991n,Sachdev}. There is some hope that
confinement-deconfinement transitions in $2+1$ dimensions may shed
light on
quantum phenomena such as spin-charge separation and the breakdown of
Fermi liquid theory in two spatial dimensions, and quantum phase
transitions in Mott insulators. One central issue is how to describe
exotic physical phenomena which do not comply with the Fermi-liquid
paradigm such as the fractional quantum Hall effect, high-$T_c$
superconductivity, and heavy fermion physics.

In this paper we will consider two different $(2+1)$-dimensional
Abelian lattice gauge theories containing a Chern-Simons (CS) term
\cite{Jackiw}.
The first of these theories is defined by a kind of lattice  Maxwell-Chern-Simons
Lagrangian.
Polyakov has demonstrated that
the compact Maxwell theory is permanently confined in $2+1$ dimensions
and does not exhibit any phase transition \cite{polyakov}.
When this theory is coupled to bosonic matter fields with an integer charge
$q>1$, a deconfinement transition occurs \cite{FradShe,sudbo2002}. The case
where the bosonic matter fields are coupled to the fundamental charge is
more controversial and at present it is not yet known if a deconfinement transition indeed
occurs \cite{KNS,Herbut}. In this paper we will
study the effect of adding a CS-term to Polyakov theory. Such a theory was
studied previously by a number of authors
\cite{cs-monopoles,Kovner}. In this paper we will use duality techniques to
obtain new {\it exact} results on a lattice Abelian compact theory with a
special type of CS-term. We
will show that a phase transition does occur as a function of the
gauge coupling for a {\it fixed}
value of the CS-coupling, while confinement of
electric charges is suppressed.

The second theory considered in this paper is a non-compact lattice
Abelian Higgs model with a CS-term.  This theory has also  been much studied
in the past by many authors \cite{read:1989,Semenoff,ReyZee,deCalan}, often
in connection with condensed matter systems. In this paper, we will
compute the critical exponents of this theory as a  function of the
CS-coupling using  Monte Carlo simulations. To our knowledge, this is
the first time Monte Carlo simulations are employed on a lattice
CS-theory.

The lattice Maxwell-Chern-Simons theory with a compact gauge field used in
this paper is defined by the partition function

\begin{eqnarray}
\label{U(1)gauge}
&Z=\int_{-\pi}^\pi\left[\prod_{i \mu} \frac{d
    A_{i\mu}}{2\pi}\right]\sum_{\{\ve{n},\ve{N}\}}
e^{-\sum_i \mathcal{L}_i}\nonumber\\
&\mathcal{L}_i=  \frac{1}{2f^2} (\deltab\times\ve{A}_i -2\pi\ve{n}_i)^2
+\frac{ig}{2}(\ve{A}_{i}-2\pi\ve{N}_i)\cdot(\deltab\times\ve{A}_{i}-2\pi\ve{n}_i),
\end{eqnarray}
where $\ve{A}_i$ is a periodic gauge field, $\ve{n}_i$ and $\ve{N}_i$ are
integer fields taking care of the compactness of the theory and $\Delta$ is the
lattice difference operator. The above Lagrangian is written in 
Villain-like form in order to allow the use of standard duality 
transformations. Note that it 
differs from the usual compact 
Maxwell-Chern-Simons Lagrangian \cite{Kovner}, where
the constraint $\deltab\times{\bf N}_i={\bf n}_i$ is expected to hold. 
Physically, this constraint leads to a complete suppression of 
magnetic monopoles (which are instantons in $2+1$ dimensions), 
since it implies $\deltab\cdot{\bf n}_i=0$.  
Relaxation of this constraint incorporates the magnetic monopoles in 
the theory, since then $\deltab\cdot{\bf n}_i=Q_i$, where $Q_i$ 
is an integer, just like in Polyakov's $(2+1)$-dimensional compact QED 
\cite{polyakov}. As we will see, removing the constraint 
$\deltab\times{\bf N}_i={\bf n}_i$ implies new interesting physics.  

A Lagrangian similar to the one in 
Eq. (\ref{U(1)gauge}) arises in
effective descriptions of  chiral spin states \cite{Wen}. Due to compactness,
Eq. (\ref{U(1)gauge}) is gauge invariant only if $g=n/2\pi$, with
$n$ integer \cite{cs-monopoles}. This is in contrast with the
non-compact Abelian theory where $g$ can be any real
number. Furthermore, Eq. (\ref{U(1)gauge}) is also invariant under the integer
gauge transformation
${\bf A}_i\to{\bf A}_i+2\pi{\bf M}_i$,
${\bf n}_i\to{\bf n}_i+\deltab\times{\bf M}_i$, and
${\bf N}_i\to{\bf N}_i+{\bf M}_i$.

Let us associate a vector field $\ve{a}$ with $\Delta\times\ve{A}-2\pi\ve{n}$ and
$\ve{b}$ with $\ve{A}-2\pi\ve{N}$ by introducing the Lagrange multiplier vector
fields $\ve{\lambda}$ and $\ve{\sigma}$ (in the following we omit
sometimes lattice subindices to simplify the notation)
\begin{eqnarray}
\widetilde{\mathcal{L}}=&\frac{1}{2f^2}\ve{a}^2+\frac{ig}{2}\ve{a}\cdot\ve{b}+i\ve{\lambda}\cdot(\deltab\times\ve{A}-2\pi\ve{n}-\ve{a})+i\ve{\sigma}\cdot(\ve{A}-2\pi\ve{N}-\ve{b}).
\end{eqnarray}
Next, we use the Poisson summation formula to replace $\ve{\lambda}$ and
$\ve{\sigma}$ by a new set of integer valued fields denoted $\ve{L}$
and $\ve{S}$
\begin{eqnarray}
\widetilde{\mathcal{L}}=&\frac{1}{2f^2}\ve{a}^2+\frac{ig}{2}\ve{a}\cdot\ve{b}+i\ve{L}\cdot(\deltab\times\ve{A}-\ve{a})+i\ve{S}\cdot(\ve{A}-\ve{b}).
\end{eqnarray}
Straightforward integration of $\ve{A}$ leads to the
constraint $\Delta\times\ve{L}=\ve{S}$ in the partition function.
Summation over the field $\ve{S}$ followed
by integration of the fields $\ve{a}$ and $\ve{b}$ yields

\begin{equation}
\label{Ltilde}
\widetilde{\mathcal{L}}=\frac{8\pi^2}{n^2f^2}(\deltab\times\ve{L})^2
-\frac{i4\pi}{n}\ve{L}\cdot(\deltab\times\ve{L}),
\end{equation}
where we have written explicitly $g=n/2\pi$. 
The crucial point to note is that when
$n=\pm 1,\pm 2$, the integer CS-term in Eq, (\ref{Ltilde})
does not contribute to the partition function.  
In such a case, the theory can be written in terms of a new integer field 
$\ve{R}=\deltab\times\ve{L}$ to obtain 
\begin{equation}
\label{loopgas}
Z={\sum_{\{\ve{R}\}}}^\prime\exp\left(-\sum_i{\frac{8\pi^2}{n^2f^2}\ve{R}_i^2}\right).
\end{equation}
where ${\sum^{\prime}_{\{\ve{R}\}}}$ denotes a constrained sum over closed vortex
loops, i.e., with the constraint $ \deltab \cdot \ve{R} =0$ implied. 
The theory defined by Eq. (\ref{loopgas}) is a theory of closed loops
interacting through  contact repulsion. This is the well known loop gas 
representation of the Abelian Higgs model with zero screening length which 
exhibits a loop-proliferation phase transition in the $3dXY$ 
universality class. There is a family of critical points given by 
$f_c^2\approx 0.33\times 8\pi^2/ n^2$, with 
the values $n=\pm 1,\pm 2$. When $n=\pm 2$, the loop gas partition 
function (\ref{loopgas}) is equivalent to  a {\it non-compact} abelian Higgs 
model in the so called ``frozen'' limit \cite{Peskin}, 
which is the basis for the  
``inverted'' $3dXY$ universality class in superconductors \cite{dasgupta1981}. 
For $n=\pm 1$, on the other hand, 
Eq. (\ref{loopgas}) corresponds to a frozen superconductor 
where the charge of the Cooper pair 
$f=2e$ is {\it fractionalized}. Therefore, 
in this case the theory in (\ref{loopgas}) can be thought as 
corresponding to a ``frozen'' superconductor with charge 
$f/2=e$.       

The constraint $ \deltab \cdot \ve{R} =0$ implies that there are no
monopoles in the spectrum of the theory with $n=\pm 1,\pm
2$. Therefore, there is no confinement of electric charges in the 
corresponding $3dXY$ phase transition. This is entirely complementary
to the theory without a CS-term: There, one has permanent confinement
and no phase transition. 

In Ref. \cite{Kovner} a continuum version of a similar model is studied
using a Hamiltonian approach via a variational analysis. There
a $XY$-like phase transition is also found, but only as a function of
$n$. In contrast with our model, the compact Maxwell-Chern-Simons theory
studied in Ref. \cite{Kovner} does not undergo any phase transition for
{\it fixed} $n$.

We next consider the Abelian Higgs model with a noncompact CS gauge
field given in the Villain approximation by
\begin{eqnarray}
  \label{action}
S=\sum_{i} \left[\frac{\beta}{2}(\deltab \theta_i -f {\bf A}_i
-2\pi {\bf n}_{i})^2
+ig{\bf A}_i\cdot\deltab \times{\bf A}_i+\lambda (\deltab\times{\bf A}_i)^2  \right],
\end{eqnarray}
where ${\bf A}_i$ is a noncompact gauge field and $\theta_i$ is a scalar phase
field. Note that here the CS coupling $g$ can be any real number,
since now we are dealing with a non-compact Abelian gauge field.
The topological defects of this model are vortex loops. An Abelian Higgs
theory with a CS term  has been proposed as an effective theory for the Laughlin
state of fractional quantum Hall systems \cite{Fradkin:1991n,read:1989}.

By introducing an auxiliary continuous field ${\bf v}_{i}$, we can
rewrite the action in Eq. (\ref{action}) as
\begin{eqnarray}
  \label{S1}
  S = \sum_i\left[\frac{1}{2\beta}{\bf v}_i^2 - i{\bf v}_i
\cdot(\deltab \theta_i -f {\bf A}_i
-2\pi {\bf n}_{i})
+ig{\bf A}_i\cdot\deltab \times{\bf A}_i+\lambda (\deltab\times{\bf A}_i)^2 \right].
\end{eqnarray}
Besides the usual gauge invariance, the action (\ref{action}) is also
invariant under the integer gauge transformation
$\theta_i\to\theta_i+2\pi l_i$, ${\bf n}_i\to{\bf n}_i+\deltab l_i$,
where $l_i$ is an integer. This suggests that a gauge fixing
on ${\bf n}_i$ would allow us to extend the limit of integration
for $\theta_i$ over the whole real line. However, since we are
dealing with integer fields, not all gauge fixings work consistently.
The widely employed gauge fixing $\deltab\cdot{\bf n}_i=0$ does not
work because it would lead to a decoupling of $\theta_i$ from ${\bf n}_i$
and consequently to a wrong loop-gas representation in the $3dXY$ limit.
Instead, we follow Ref. \cite{Kleinert} and fix the axial gauge $n_3=0$.
Within this gauge fixing we can integrate out $\theta_i\in(-\infty,\infty)$
in Eq. (\ref{S1}) to obtain the constraint $\deltab\cdot{\bf v}_i=0$.
This constraint is solved by introducing a new vector field
${\bf h}_i$ such that ${\bf v}_i=\deltab\times{\bf h}_i$. Next
we perform a partial summation in the term
$2\pi i{\bf n}_i\cdot( \deltab\times{\bf h}_i)$ to give
$2\pi i{\bf h}_i\cdot(\deltab\times{\bf n}_i)$ and identify
the integer field ${\bf m}_i=\deltab\times{\bf n}_i$ as the
vortex field. The end result is
\begin{eqnarray}
\label{Sh}
S=\sum_i\left[\frac{1}{2\beta}(\deltab\times{\bf h}_i)^2
-i f(\deltab\times{\bf h}_i)\cdot{\bf A}_i
+2\pi i ~{\bf h}_i\cdot{\bf m}_i
+ig{\bf A}_i\cdot\deltab \times{\bf A}_i+\lambda (\deltab\times{\bf A}_i)^2
\right],
\end{eqnarray}
where the constraint $\deltab\cdot{\bf m}_i=0$ is understood.
Setting $\lambda=0$ and integrating out ${\bf A}_i$ yields
a well known self-duality \cite{ReyZee}
at $\beta\to\infty$.

After integrating out ${\bf A}_i$ and ${\bf h}_i$ and performing
a lattice Fourier transform, we obtain

\begin{eqnarray}
\label{intphichi}
&S =\sum_q m_q^\mu V_q^{\mu\nu}m_{-q}^\nu,\nonumber \\
&V^{\mu\nu}_{q} = \frac{2\beta\pi^2}{(\varphi Q_qQ_{-q} + \chi)^2 + Q_qQ_{-q}}
\left[(\varphi^2Q_qQ_{-q}+\varphi\chi +1)\delta_{\mu\nu}-\frac{i\chi\epsilon_{\mu\nu\lambda}Q_{-q}^\lambda}{Q_qQ_{-q}}\right]
\end{eqnarray}
where $Q_qQ_{-q}\equiv Q_q^\mu Q_{-q}^\mu$, with
$Q_{q}^\mu=e^{iq^\mu/2} -e^{-iq^\mu/2}$
being the Fourier representation of the symmetrized difference operator,
and we have defined $\varphi = \lambda/g$ and $\chi = \beta f^2/(2g)$.
$V^{\mu\nu}_{q}$ is the Fourier transform of the vortex-vortex interaction
tensor. In the two limits $g \rightarrow\infty$ and $f = 0$ (\ref{intphichi})
reduces to the interaction potential of vortices in the 3dXY model
$V_{\rm 3dxy}^{\mu\nu} = (2\pi^2\beta/Q_qQ_{-q}) \delta^{\mu\nu}$.

We next study the model Eq. (\ref{intphichi}) using Monte Carlo 
(MC) simulations. Having real-valued  vortex variables requires working in real-space. 
However, the action defined by Eq. (\ref{intphichi}) is complex in real-space, $S=S_R+iS_I$, 
leading to a complex transition-probability between various vortex-configurations
in the Metropolis algorithm. This is analogous to 
the ``sign problem'' in quantum Monte Carlo simulations, which is often treated by using the 
absolute value of the probability $\rho$ in the Metropolis update and sampling expectation values by 
$\langle \mathcal{O}\rangle = \langle\mathcal{O}s\rangle_{|\rho|}/\langle s\rangle_{|\rho|} $, 
where $s$ is the sign of the probability \cite{vonderLinden1992}. The expectation 
value of an operator $\mathcal{O}$ is in our case defined schematically by
\begin{equation}
\langle \mathcal{O}\rangle_S
=\frac{\int\mathcal{D}\Psi \mathcal{O}e^{S_R+iS_I}}{\int\mathcal{D}\Psi e^{S_R+iS_I}}
\end{equation}
where $\Psi$ denotes the complete set of eigenstates given by
$S$. Since the real part of the action $S_R$ defines the same set of
eigenstates, i.e. closed vortex loops, we may express
$\langle\mathcal{O}\rangle_S$ in terms of expectation values
$\langle\rangle_{S_R}$ 
\begin{eqnarray}
\label{operator}
\langle \mathcal{O}\rangle_S=\frac{\int\mathcal{D}\Psi (\mathcal{O}e^{iS_I})e^{S_R}/Z_{S_R}}{\int\mathcal{D}\Psi (e^{iS_I})e^{S_R}/Z_{S_R}}=\frac{\langle \mathcal{O}e^{iS_I}\rangle_{S_R}}{\langle e^{iS_I}\rangle_{S_R}},
\end{eqnarray}
where $Z_{S_R}$ is the partition function for the system defined by the real part of the action.
The MC simulations are performed using $e^{S_R}$ as the Boltzmann weight and expectation values
are calculated from Eq. (\ref{operator}).

The system size $L$ and the lattice constant define the only
length scales of the system. Critical properties are governed by the
long range physics i.e. the $q \to 0$ limit. In this limit the real part
of the potential (\ref{intphichi}) has an effective screening length
$\lambda_{\rm eff}^{-2} \sim \chi^2/(2\varphi\chi + 1)$.
When this screening length is of the order of $L/2$, the critical
behavior will probably be a crossover to the 3dXY model with an infinite screening
length. On the other hand, when $\lambda_{\rm eff} < a$ the critical behavior
will experience a crossover towards a system with steric repulsion. Hence, MC
simulations on finite lattices can only provide true critical exponents
in a limited region of coupling space where $a\ll\lambda_{\rm{eff}}\ll
L/2$.

\begin{figure}[htbp]
\centerline{\scalebox{0.6}{\rotatebox{0.0}{\includegraphics{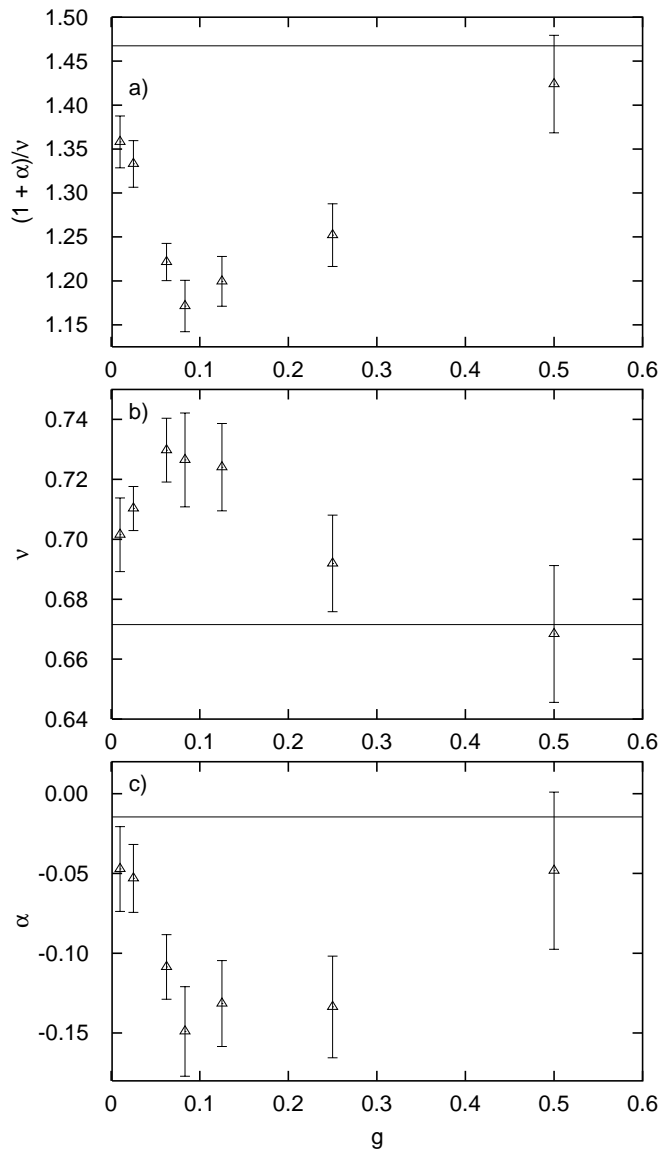}}}}
\caption{\label{scalplot} a) $(1+\alpha)/\nu$ from finite size scaling
of $M_3$. The solid line denotes the well known $3dXY$
values and is included here for the sake of comparison. 
b) Critical exponent $\nu$ computed directly from the width of
$M_3$. c) $\alpha$ from combining $M_3$ results for
$(1+\alpha)/\nu$ and $1/\nu$. }
\end{figure}

We have performed MC simulations on the model defined by
\begin{equation}
\label{finalaction}
Z={\sum_{\{m_\mu(r_i)\}}}^\prime  \exp{\Bigl[-\sum_{i,j}m_\mu(r_i)V^{\mu\nu}(r_i-r_j)m_\nu(r_j)\Bigr]},
\end{equation}
where the potential $V^{\mu\nu}(r)$ is the inverse Fourier transform of
Eq. (\ref{intphichi}). A MC move is an attempt to insert a
unitary closed vortex loop of random orientation, and the move is accepted or
rejected according to the standard Metropolis algorithm using the real
part of the action.
One sweep consists of traversing the lattice performing one MC move for each
lattice site. The system with size $L\times L\times L$ has periodic
boundary conditions. Simulations with up to $10^6$ sweeps per coupling
constant combined with Ferrenberg-Swendsen multihistogram reweighting
are used to produce finite size scaling (FSS) plots. System sizes
$L \in [4,6,8,10,12,14,16,20,24]$ have been used.

To investigate the critical properties of the model, we compute critical
exponents $\alpha$ and $\nu$. Recently we have proposed a FSS method based
on the third moment $M_3$ of the action $S$ \cite{sudbo2002},
$M_3 = <(S - \langle S \rangle)^3>_S$,
for which the asymptotically correct behavior is reached for accessibly small system
sizes. The peak
to peak value of this quantity scales with system size as $L^{(1+\alpha)/\nu}$ and
the width between the peaks scales as $L^{-1/\nu}$. In this way one can resolve
$\alpha$ and $\nu$ independently from one measurement without
invoking hyperscaling \cite{sudbo2002}.

The denominator $|\langle e^{iS_I}\rangle_{S_R}|$ decreases with system size and
coupling strength, contaminating expectation values Eq. (\ref{operator}).
Hence, as the strength of the CS term $g$ increases, smaller system sizes are
accessible for the FSS. Together with the limitations on the
effective screening length $\lambda_{\rm eff}$, this restricts the range in
coupling space for which critical exponents can be calculated. We compute
exponents along a line in coupling space corresponding to fixing $f=1$,
$\lambda=\frac{1}{4}$, and tuning $g$ so that Eq. (\ref{operator}) is meaningful
and crossover effects are absent. Along this line exponents can only be
extracted from $M_3$-analysis for $g\le 0.5$. We simulate system sizes $L=4,
6, 8,\dots$ where the maximum system size varies from $L=12$ for $g=0.5$ to at
least $L=24$ for $g=0.01$. The results of the FSS analysis are
shown in Fig. \ref{scalplot}. We find non-universal exponents $\alpha$ and
$\nu$ approaching 3dXY values for large $g$ and when $g\to 0$.
Continuously varying critical exponents are a consequence of the
vanishing of the renormalization group (RG)
$\beta$-function of the CS-coupling \cite{Semenoff,Ferretti}.
They are associated with a marginal operator, which in this case
is just the CS-term. Critical exponents  for the Abelian
Higgs model with a CS-term have been obtained  previously
using RG  methods \cite{deCalan,Ferretti}. Here we provide for the first time a
MC calculation of these exponents.
Due to the technical difficulties explained in Ref. \cite{deCalan},
the RG calculations are more reliable for a generalized model with
$N/2$ complex field components where $N$ is large enough. Although
our results are for $N=1$, it is useful to compare at least qualitatively
the results obtained here with those obtained from the $1/N$-expansion.
In this case the critical exponent $\nu$ is given at
order $1/N$ by \cite{Ferretti}

\begin{equation}
\label{nu}
\nu=1-\frac{96}{\pi^2N}\left[1-\frac{8}{9}\frac{\bar{g}^2(\bar{g}^2
+4)}{(\bar{g}^2+1)^2}\right],
\end{equation}
where we have defined $\bar{g}=4g/\pi$. The critical exponent
$\nu$ as given by Eq. (\ref{nu}) exhibits a similar qualitative
behavior in comparison with panel (b) in Fig. \ref{scalplot}, having
a maximum at some value of $g$. For smaller values of $N$, on the
other hand, the RG treatment is in poor agreement with our numerical
results, both from qualitative and quantitative points of view \cite{deCalan}.

In summary, we have shown in two different gauge theories that the presence of a CS
term  changes dramatically their behavior. In both cases the analysis was entirely
non-perturbative and based on duality arguments.

\acknowledgments
The authors would like to thank H. Kleinert for discussions.
This work received financial support from the Norwegian University of Science and
Technology through the Norwegian High Performance Computing Program (NOTUR),
from  the Research Council of Norway, Grant Nos. 157798/432,
158518/431, 158547/431 (A.S), and from DFG Priority
Program SPP 1116 (F.S.N.).



\begin{thebibliography}{100}

\bibitem{polyakov} \Name{Polyakov, A.M} \REVIEW{Nucl. Phys. B }{120}{1977}{429}.

\bibitem{Nagaosa} \Name{Lee, P.A. \and Nagaosa, N.} \REVIEW{Phys. Rev. B}{46}{1992}{5621};
\Name{Mudry, C. \and Fradkin, E.} \REVIEW{Phys. Rev. B}{50}{1994}{11409};
\Name{Balents, L., Fisher, M.P.A. \and Nayak, C.}
\REVIEW{Phys. Rev. B}{61}{2000}{6307}; \Name{Ichinose, I., Matsui, T., \and Onoda, M.}
\REVIEW{Phys. Rev. B}{64}{2001}{104516}.

\bibitem{Fradkin:1991n} \Name{Fradkin, E.}
\Book{Field Theories of Condensed Matter Systems}
\Publ{Addison-Wesley, Reading, MA} \Year{1991}, and references therein.

\bibitem{Sachdev} \Name{Read, N. \and Sachdev, S.} \REVIEW{Phys. Rev. Lett.}{66}{1991}{1773};
\Name{Sachdev, S. \and Read, N.} \REVIEW{Int. J. Mod. Phys. B}{5}{1991}{219};
\Name{Sachdev, S. \and Park, K.} \REVIEW{Ann. Phys. (N.Y.)}{298}{2002}{58};
\Name{Sachdev, S.} \REVIEW{Rev. Mod. Phys.}{75}{2003}{913}.

\bibitem{Jackiw} \Name{Deser, S., Jackiw, R. \and Templeton, S.}
\REVIEW{Ann. Phys. (N.Y.)}{140}{1982}{372}.

\bibitem{FradShe} \Name{Fradkin, E. \and Shenker, S.H.} \REVIEW{Phys. Rev. D}{19}{1979}{3682}.

\bibitem{sudbo2002} \Name{Sudb{\o}, A., Sm{\o}rgrav, E.,  Smiseth, J., Nogueira, F.S.
\and Hove, J.} \REVIEW{Phys. Rev. Lett.}{89}{2002}{226403};
\Name{Smiseth, J., Sm{\o}rgrav, E., Nogueira, F.S., Hove, J. \and Sudb{\o}, A.}
\REVIEW{Phys. Rev. B}{67}{2003}{205104}.

\bibitem{KNS} \Name{Kleinert, H., Nogueira, F.S. \and Sudb{\o}, A.}
\REVIEW{Phys. Rev. Lett.}{88}{2002}{232001}; \REVIEW{Nucl. Phys. B}{666}{2003}{361}.

\bibitem{Herbut} \Name{Herbut, I.F. \and Seradjeh, B.H.} \REVIEW{Phys.Rev.Lett.}{91}{2003}{171601};
\Name{Herbut, I.F., Seradjeh, B.H., Sachdev, S. \and Murthy, G.}
\REVIEW{Physical Review B}{68}{2003}{195110}.

\bibitem{cs-monopoles} \Name{d'Hoker, E. \and Vinet, L.} \REVIEW{Ann. Phys. (N.Y.)}{162}{1985}{413};
\Name{Pisarski, R.D} \REVIEW{Phys. Rev. D}{34}{1986}{3851};
\Name{Affleck, I., Harvey, J., Palla, L. \and Semenoff, G.} \REVIEW{Nucl. Phys. B}{238}{1999}{575};
\Name{Fradkin, E. \and Schaposnik, F. A.} \REVIEW{Phys. Rev. Lett.}{66}{1991}{276}.

\bibitem{Kovner} \Name{Kogan, I. \and Kovner, A.} \REVIEW{Phys. Rev. D}{53}{1996}{4510}.

\bibitem{read:1989} \Name{Read, N.} \REVIEW{Phys. Rev. Lett.}{62}{1989}{86};
\Name{Zhang, S.C., Hansson, T. \and Kivelson, S.} \REVIEW{Phys. Rev. Lett.}{62}{1989}{82}.

\bibitem{Semenoff} \Name{Semenoff, G.W., Sodano, P., \and Wu, Y.-S.}
\REVIEW{Phys. Rev. Lett.}{62}{1989}{715}.


\bibitem{deCalan} \Name{de Calan, C., Malbouisson, A.P.C., Nogueira, F.S. \and Svaiter, N.F.}
\REVIEW{Phys. Rev. B}{59}{1999}{554};
\Name{Kleinert, H. \and Nogueira, F.S.} \REVIEW{J. Phys. Stud.}{5}{2001}{327}.

\bibitem{ReyZee} \Name{Rey, S.J. \and Zee, A.} \REVIEW{Nucl. Phys. B}{352}{1991}{897};
\Name{Diamantini, M.C., Sodano, P. \and Trugenberger, C.A.}
\REVIEW{Phys. Rev. Lett.}{71}{1993}{1969}; \REVIEW{Phys. Rev. Lett.}{75}{1995}{3517}.

\bibitem{Wen} \Name{Wen, X.G., Wilczeck, F. \and Zee, A.}
\REVIEW{Phys. Rev. B}{39}{1989}{11413}.


\bibitem{Peskin} \Name{Peskin, M} \REVIEW{Ann. Phys. (N.Y.)}{113}{1978}{122}.

\bibitem{dasgupta1981} \Name{Dasgupta, C. \and Halperin, B.I.}
\REVIEW{Phys. Rev. Lett.}{47}{1981}{1556}.


\bibitem{Kleinert} \Name{Kleinert, H.} \Book{Gauge Fields in Condensed
Matter, vol. 1} \Publ{World Scientific, Singapore} \Year{1989}, page 561,
readable at
http://www.physik.fu-berlin.de/\~{}kleinert/re.html\#b1.

\bibitem{vonderLinden1992} \Name{von der Linden, W.} \REVIEW{Phys. Rep.}{220}{1992}{53}.


\bibitem{Ferretti} \Name{Ferretti, G. \and Rajeev, S.G.} \REVIEW{Mod. Phys. Lett. A}{7}{1992}{2087}.


\end{thebibliography}
\end{document}